%% file: main.tex
\documentclass[conference]{IEEEtran}
\pdfoutput=1
\IEEEoverridecommandlockouts
\usepackage{pdfpages}
\usepackage{hyperref}
\usepackage[absolute]{textpos}
\usepackage{cite}
\usepackage{amsmath,amssymb,amsfonts}
\usepackage{algorithmic}
\usepackage{graphicx}
\usepackage{textcomp}
\usepackage{xcolor}
\usepackage{comment}
\usepackage{tabularx} 
\usepackage{booktabs}
\usepackage{multirow}
\usepackage{hyperref}
\usepackage{wrapfig}

\renewcommand{\baselinestretch}{.96}

\begin{document}
\begin{textblock}{24}(1,0.2)
    \noindent\tiny  This paper is a preprint; it has been published in 2021 IEEE International Conference on Cyber Security and Resilience (CSR), Workshop on Data Science for Cyber Security (DS4CS @ IEEE CSR), Rhodes, Greece, July  \\
    \textbf{IEEE copyright notice} \textcopyright 2021 IEEE. Personal use of this material is permitted. Permission from IEEE must be obtained for all other uses, in any current or future media, including reprinting/republishing this material for advertising or promotional purposes,\\ creating new collective works, for resale or redistribution to servers or lists, or reuse of any copyrighted component of this work in other works.
    \end{textblock}

\title{{Social Media Monitoring for IoT Cyber-Threats}\\
    \thanks{
        \protect\begin{wrapfigure}[3]{l}{1cm}
        \protect\raisebox{-12.5pt}[0pt][7pt]{\protect\includegraphics[height=.9cm]{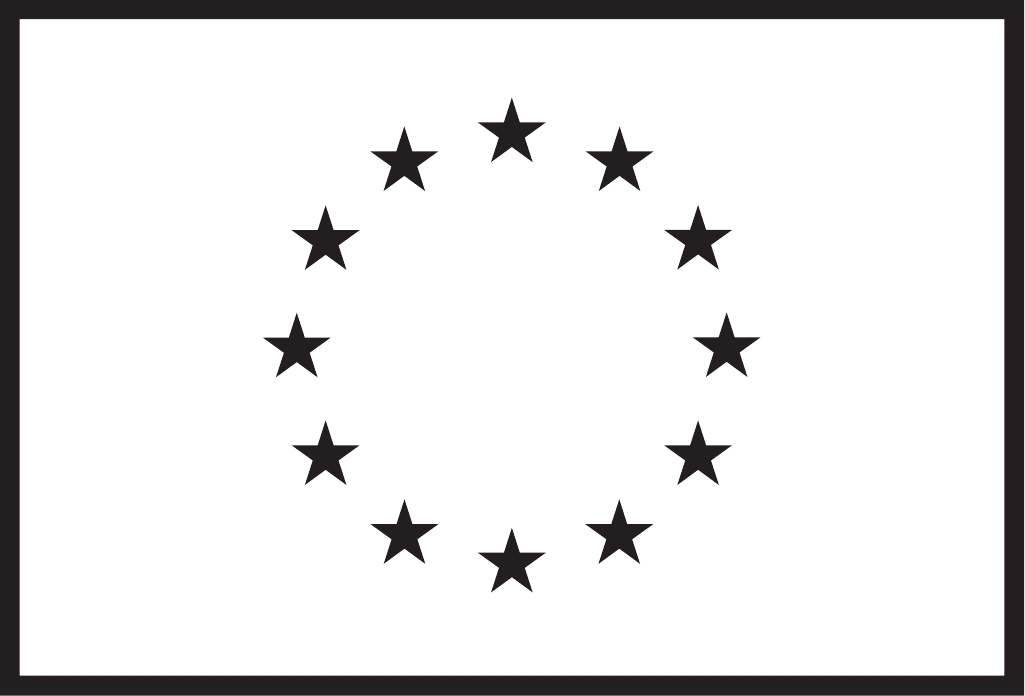}}
        \protect\end{wrapfigure}
            This work has received funding from the European Union's Horizon 2020 research and innovation programme under grant agreements no. 786698 and no.833673. The work reflects only the authors' view and the Agency is not responsible for any use that may be made of the information it contains.
    }
}

\author{\IEEEauthorblockN{Sofia Alevizopoulou\IEEEauthorrefmark{1},
Paris Koloveas\IEEEauthorrefmark{2}, Christos Tryfonopoulos,\IEEEauthorrefmark{3} and
Paraskevi Raftopoulou\IEEEauthorrefmark{4}}
\IEEEauthorblockA{Dept. of Informatics \& Telecommunications,
University of the Peloponnese, GR22131 Tripolis, Greece\\
email: \IEEEauthorrefmark{1}dsc17002@uop.gr,
\IEEEauthorrefmark{2}pkoloveas@uop.gr,
\IEEEauthorrefmark{3}trifon@uop.gr,
\IEEEauthorrefmark{4}praftop@uop.gr}}

\maketitle
\input{sec0_abstract/abstract.tex}
\input{sec1_introduction/introduction.tex}
\input{sec2/sec2.tex}

\input{sec3/sec3.tex}
\input{sec4/sec4.tex}

\input{sec5/sec5.tex}
\input{sec7/sec7.tex}

\bibliographystyle{unsrt}
\bibliography{references}

\end{document}

%% file: sec0_abstract/abstract.tex
\begin{abstract}
The rapid development of IoT applications and their use in various fields of everyday life has resulted in an escalated number of different possible cyber-threats, and has consequently raised the need of securing IoT devices. Collecting Cyber-Threat Intelligence (e.g., zero-day vulnerabilities or trending exploits) from various online sources and utilizing it to proactively secure IoT systems or prepare mitigation scenarios has proven to be a promising direction. In this work, we focus on social media monitoring and investigate real-time Cyber-Threat Intelligence detection from the Twitter stream. Initially, we compare and extensively evaluate six different machine-learning based classification alternatives trained with vulnerability descriptions and tested with real-world data from the Twitter stream to identify the best-fitting solution. Subsequently, based on our findings, we propose a novel social media monitoring system tailored to the IoT domain; the system allows users to identify recent/trending vulnerabilities and exploits on IoT devices. Finally, to aid research on the field and support the reproducibility of our results we publicly release all annotated datasets created during this process.
\end{abstract}


%% file: sec1_introduction/introduction.tex
\vspace{-1mm}
\section{Introduction}\label{sec:intro}

The term IoT typically refers to a number of interrelated computing devices (e.g., sensors, smart devices and similar lightweight computing systems) that send and receive data over the Internet~\cite{GSMA}. In recent years, IoT technology has been widely spread and has attracted great interest both in industry and in academia; IoT devices are expected to spread even more in the coming years, aiming at major improvements in several fields of everyday life. This tremendous growth in the number of IoT devices has resulted in a continuous exchange of information among devices and users over the Internet. This in turn has raised many security issues and challenges to be addressed since both the exchanged data and the IoT devices themselves are exposed to vast numbers of attacks of varying impact and sophistication. 

In order to mitigate such threats, relevant Cyber-Threat Intelligence (CTI) may be collected from various online sources such as forums, blogs, social media and marketplaces, and appropriately utilized to proactively safeguard IoT systems and ensure their secure and efficient operation. This is particularly important in an IoT device given the difficulty to apply patches and other security updates to it, the large number of legacy systems, and the inability of the devices to employ any existing sophisticated mechanism to recognize new threats.
Among the several CTI sources, social media is one of the most common and dynamic alternatives for capturing new threats and attacks; people's familiarity with technology, the increased awareness for cyber-security issues in the general public and the timeliness of information delivery in social media has contributed to an unprecedented increase in the number of cyber-threat related topics being posted in these platforms, which in turn has produced huge amounts of relevant and (potentially useful) CTI data. Twitter is one of the most popular social media platforms meant for both casual and technical social interactions that offers publicly available data, a constant stream of up-to-date information around many topics (including the latest cyber-security issues), and extensive accessibility via appropriate APIs; due to its popularity and accessibility, numerous Twitter monitoring systems on various topics have emerged~\cite{Sauerwein, Syed, Sapienza, Mittal, Alves, Sabottke, Ba-Dung}.

In this work, we investigate the problem of identifying CTI-related tweets from the Twitter stream; to do so we (i) create and appropriately annotate a dataset that contains security tweets with/without IoT-related CTI, (ii) extensively evaluate six different machine-learning (ML) based classification algorithms with the aforementioned datasets to identify the best fitting approach, and (iii) develop and deploy a novel social media monitoring scheme that allows users to identify tweets with recent/trending vulnerabilities and exploits on IoT devices. More specifically, to create the test base for our experiments we obtain IoT security-related tweets from the Twitter API and assign them to cyber-security experts that annotate them based on whether they contain CTI or not. Subsequently, we experiment with six traditional ML algorithms, (Logistic Regression, Multinomial Naive Bayes, Decision Trees, K-Nearest Neighbors, Support Vector Machines and Random Forest) that are typically employed in Twitter classification tasks to build models that allow the automatic identification of CTI-related tweets from Twitter streams. Notice that classifying IoT security tweets to CTI and non-CTI related is a non-trivial task, very different from other tweet classification tasks in the literature due to the very specific nature of the task and the highly common vocabulary between the two tweet classes. Finally, based on our findings on the best performing algorithm and setup, we design, develop, deploy and evaluate a novel monitoring system to identify CTI-related tweets in real-time. Based on the above, our contributions are threefold.
\begin{itemize}
  \item This is the \textit{first work} that studies the effectiveness of several popular classification alternatives for \textit{IoT-related CTI detection on social media} content. Notice that this is a particularly difficult task given the short textual content of the Twitter platform, the wide vocabulary introduced by the vast volume of IoT devices and models, and the common technical vocabulary between casual security discussions/announcements and CTI posts. Based on our findings we develop and evaluate a \textit{novel} monitoring system to identify CTI-related tweets in real-time.
  \item We \textit{push the knowledge boundaries} on the topic since the obtained results from our experiments \textit{contradict} other findings in more general contexts. Our findings designate \textit{Random Forest} as the algorithm of choice for the task at hand, in contrast to findings in other similar tasks (e.g., vulnerability classification, CTI categorization) that identified mostly SVMs as the model of preference for content classification~\cite{Blinowski, Shuai, Queiroz}.
  \item We openly release a \textit{new dataset}~\cite{myrepo} comprised of security-related tweets annotated according to whether they contain IoT CTI; this is expected to aid research in the area of security-oriented social media content classification and support the reproducibility of our results.
\end{itemize}

The rest of this paper is organized as follows. Section~\ref{sec:relatedWork} studies the related work on the CTI domain, reviews existing classification models used to classify vulnerabilities, and discusses existing monitoring systems for the Twitter stream. Section~\ref{sec:datasets} describes the data preprocessing phase and presents the construction of the training, evaluation and testing datasets created for the evaluation of the different classification methods. Section~\ref{sec:classification}, outlines the ML algorithms evaluated in order to find the best model for our set-up, and discusses the selected classification model used in the developed monitoring system, while Section~\ref{sec:monitoring} reports the results concerning the experimental evaluation of the monitoring system. Finally, Section~\ref{sec:conclusion} concludes the paper and outlines future research.

%% file: sec2/sec2.tex
\section{Related Work}\label{sec:relatedWork}

The rapid development of IoT applications and their use in various fields of everyday life has been followed by an escalated number of possible threats. This in turn, has resulted in many studies related to threat intelligence as a mean to deal with the security issues arisen.
In what follows, we present the state-of-the-art research related to vulnerabilities classification and CTI, focusing on social media monitoring.

\subsection{Vulnerabilities classification}
As the number and the variety of vulnerabilities have gradually increased, their management and analysis has become a critical issue. Common Vulnerabilities and Exposures (CVE) identifier, launched in 1999 by MITRE~\cite{cve-website}, identifies, defines, and catalogs publicly disclosed cyber-security (CS) software and firmware vulnerabilities; CVE lists the already known vulnerabilities and exposures with a unique ID, followed also by a brief description and references to related vulnerability reports and advisories, allowing this way security administrators to access technical information concerning a specific threat across multiple CVEs. Apparently, when a vulnerability can be classified, the risk of a system being attacked (and damaged) is reduced, since the vulnerability can be managed more efficiently. 

The work in~\cite{Blinowski} is a hand-on analysis on NVD~\cite{NVD} entries with hardware CPE and proposes the grouping of the records into seven categories (i.e., home, car, service providers, etc.) from the perspective of attack. The annotated records are then used in a SVM algorithm trained to classify ``new'' records that describe vulnerabilities of IoT devices used for different applications. 
An automatic vulnerability classification framework, based on conditions that activate vulnerabilities, and different ML techniques to construct a classifier with the highest F-measure are proposed and tested in~\cite{Davari}. In~\cite{Shuai}, a novel ML method based on LDA model and SVM for automatic classification of vulnerabilities is presented; the experimental results showed that it obtains high classification accuracy and efficiency. Similarly, a SVM classifier that classifies software vulnerabilities achieving 94\% accuracy is presented in~\cite{Queiroz}. A Naïve Bayes (NB) classifier, at which the textual description of the bug report was used to extract textual information, is presented in~\cite{Dumidu}. 
The work in~\cite{Huang} proposes the automatic vulnerability classification model TFI-DNN, which is based on term frequency-reverse document frequency (TF-IDF), information gain (IG), and deep neural networks (DNN). TFI-DNN was trained and tested using vulnerability data from the NVD and the results showed that this model effectively improves the performance of vulnerability classification. 
An automatic classification model that classifies vulnerabilities based on the vulnerability description is presented in~\cite{Gawron17automatic}, where two different approaches were tested and the experimental results showed that a neural network (NN) model is superior to a NB classifier. 
In~\cite{harer2018automated}, a data-driven approach for vulnerability detection, at which features of deep model learning with tree-based models were combined, is proposed. The experimental results showed that this combination had better performance than a pure deep NN model or a traditional Random Forest classifier.

\subsection {Cyber-Threat Intelligence}
Cyber-Threat Intelligence gathering is a subdomain of cyber-security that concerns the collection of intelligence disciplines via Internet; a full CTI cycle contains the mechanism of data acquisition, data process, knowledge identification and sharing. All these processes are implemented inside a monitoring system, which collects related data from several intelligence sources after being trained over specific data in order to acquire as much related data as possible. CTI contains research works about ongoing threats, malware information, related information crawled from the web and the social media, vulnerabilities alerts or exposures, etc. Recent works~\cite{CTI-crawler, intime} propose a ML-based crawler for harvesting the clear, social, and dark web to find out IoT-related CTI data, where an SVM classifier has been used to direct the crawl towards topically relevant websites.

Social media, and specifically Twitter constitutes one of the most popular CTI sources that has been used to collect knowledge about vulnerabilities, threats, and incidents. In~\cite{Sauerwein}, a monitoring system that collects and analyses tweets related to security vulnerabilities is developed. This work concludes that the information shared on Twitter appears earlier than any official announcement and thus, the use of this information can improve the reaction to newly discovered vulnerabilities. 
The work in~\cite{Syed} analyses the content of collected tweets in order to identify the major categories in software vulnerabilities and attempts to understand the factors that impact the retweeting of software vulnerability posts. A system that detects security threats from Twitter stream data by utilizing Twitter and dark web data in order to generate appropriate warnings is introduced in~\cite{Sapienza}; the proposed method was shown capable of predicting DDoS attacks. In~\cite{Mittal}, a system that discovers and analyses CTI on Twitter data and alert accordingly the users is proposed. 
In a similar spirit,~\cite{Alves} presents a Twitter streaming-based threat monitor that generates a continuously updated summary of the threat landscape, whereas~\cite{Sabottke} designs a Twitter-based exploit detector used for early detection of real-world exploits. In~\cite{Ba-Dung}, a monitoring system is developed; it gathers CTI data from Twitter using a novelty detection model and is trained with the threat descriptions from public repositories such as CVE. The classification results suggest that cyber-threat relevant tweets do not often include the CVE identifier and for that reason, analyzing them and finding the related CVE identifier could provide further useful information.

Contrary to previous works, our approach is the first to detect IoT vulnerabilities on Twitter, also being trained with (balanced) CVE records and tested with actual Twitter data (all used datasets will be available on Github). In addition, our evaluation process highlighted the Random Forest classifier, contrary to previous studies, as the most suitable algorithm to be used in the given social media monitoring system.

%% file: sec3/sec3.tex
\section{Datasets}\label{sec:datasets}

\begin{table*} [t]
 \centering
 \caption{Example tweets related and unrelated to IoT vulnerabilities}\label{table:related-not-related-tweets-test}
 \vspace{-1mm}
    \begin{tabularx}{\textwidth}{XX}
        \toprule
        \textbf{\centering Related to IoT vulnerabilities}  & \textbf{Unrelated to IoT vulnerabilities} \\
        \midrule 
        \textit{CVE-2021-22882 UniFi Protect before v1.17.1 allows an attacker to use spoofed cameras to perform a denial-of-service attack that may cause the UniFi Protect controller to crash.} & \textit{CVE-2021-20987 A denial of service and memory corruption vulnerability was found in Hilscher EtherNet/IP Core V2 prior to V2.13.0.21that may lead to code injection through network or make devices crash without recovery.} \\ 
        \midrule
        \textit{Misconfigured Baby Monitors Allow Unauthorized Viewing \#CloudSecurity \#IoT \#Vulnerabilities \#InfoSec \#MobileSecurity}  &   \textit{Could Blackberry have a real chance in IoT? \#IoT}\\
        \bottomrule
    \end{tabularx}
\end{table*}

\begin{figure}[t]
\centerline{\includegraphics[width=0.5\textwidth]{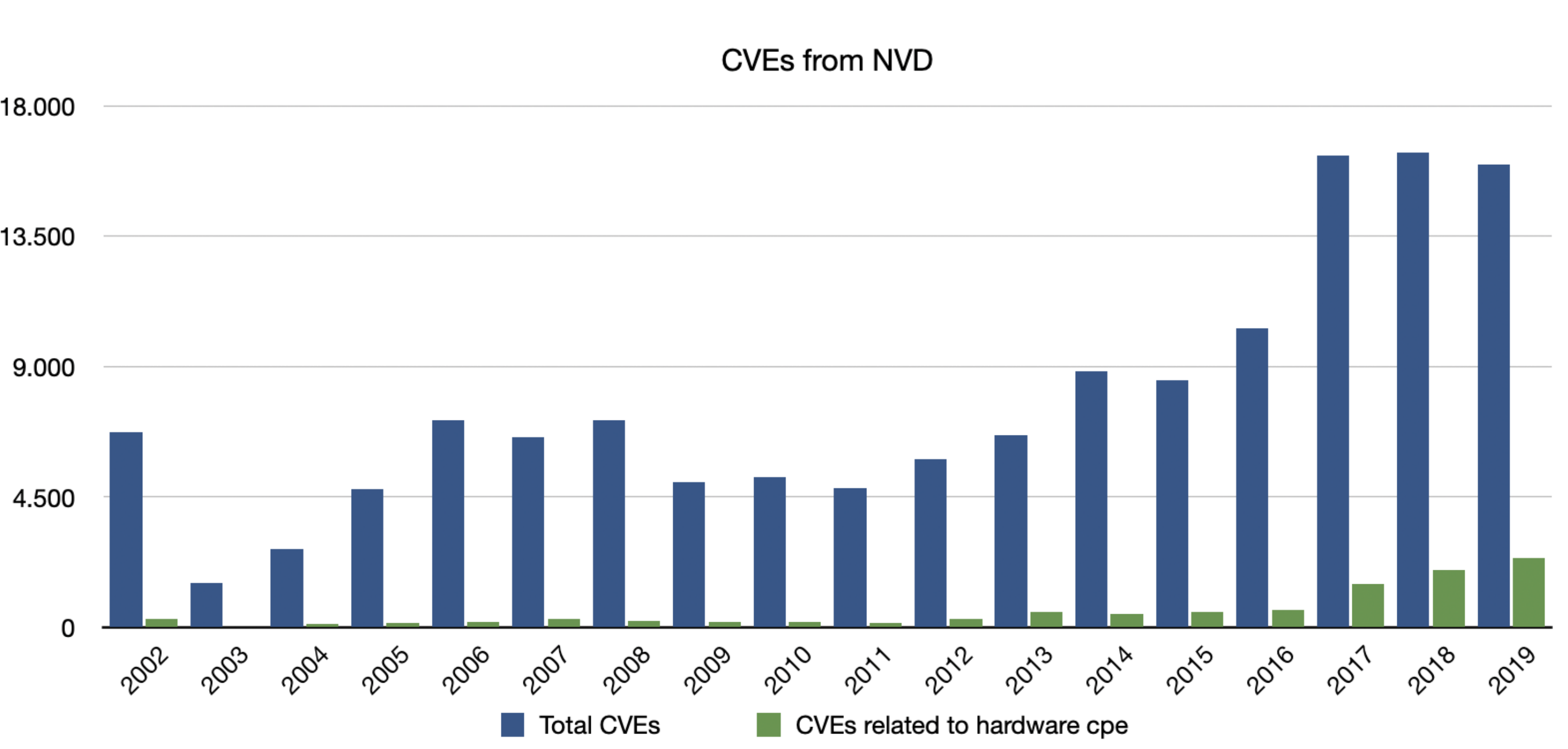}}
\vspace{-2mm}
\caption{CVEs from NVD list}\label{fig:cves_2002_2019}
\vspace{-5mm}
\end{figure}

\noindent\textbf{Training and validation datasets.} We have created a large dataset (concerning the 2002--2019 time period) with thousands of CVEs extracted from the NVD database. 
As we would like to build a classifier that can identify posts related and unrelated to IoT vulnerabilities, we filtered the collected CVEs. 
For the filtering mechanism, we were based on the fact that when a CVE has at least one hardware CPE descriptor, then this record refers to devices or systems that can potentially be a component of the perception or network layer of an IoT device; these CVEs were defined as CVEs related to IoT vulnerabilities. The remaining CVEs, with application or software CPE descriptors, were considered as unrelated to vulnerabilities of IoT devices. 
From the 140,380 CVE records, only the 9,941 records are related with at least one hardware CPE (Figure~\ref{fig:cves_2002_2019}). Note that duplicated and rejected CVEs have been removed. 

Upon a close examination on the CVE records, we observed that words and phrases such as ``use'', ``remote attacker'', ``allows remote'', ``attackers cause'', ``denial service'', etc., as well as the substring ``CVE-****-****'', are common in both related and unrelated to IoT vulnerabilities records. %Focusing on the CVE tags and 
Taking into account that cyber-threat relevant tweets do not often include a CVE identifier, as well as that we would like to build a classification model that performs well regardless the existence of a CVE tag in the tweet, we tested two different versions of the CVE dataset: (i) we removed the CVE tag from all records of the dataset and (ii) we let the context of the CVE entries unmodified. The two dataset versions were then tested using traditional machine learning algorithms in order to find out which classification model and which dataset version when combined have the best classification results. We expected that the model being trained with the no CVE-tagged data would be more generic and thus, would fit better with the data collected from Twitter.

We shuffled the above described CVE records and split them again in order to create the training and validation (used for hyper-parameter tuning) datasets, which are a subset of the initial datasets consisting respectively of 8,924 and 4,396 CVE records. Both datasets were balanced in order to train an unbiased classifier.

\noindent\textbf{Testing dataset.} Twitter provides access to its data through the application programming interface\footnote{\url{https://developer.twitter.com/en/docs/twitter-api}} (API) endpoints that can be used to learn from and engage with the conversations on Twitter. An application can be created and registered on Twitter's website, where an access key, that authenticates the user of the API, is provided, whereas rate limits are applied to endpoints based on which authentication method is used. %\footnote{https://developer.twitter.com/en/apply-for-access.html}. 

\textit{Get Tweet timelines API} and \textit{Twitter Streaming API}
are two of the most popular APIs. \textit{Get Tweet timelines API} returns a dataset of tweets that have already been posted; it involves polling Twitter’s data from specific Twitter timelines through a search based on some specific search criteria including: user-id, screen-name, count, etc. \textit{Twitter Streaming API} returns a dataset with tweets that happen in near real-time; it involves polling Twitter's data through a search based on some criteria but also maintains the connection open in order to return data whenever there will be a tweet that matches the criteria. Using the Streaming API, users request tweets that match specific search criteria such as usernames, keywords, locations, etc. 
Both APIs are widely used for data acquisition in several monitoring systems and each API client responds with a JSON object
%\footnote{https://developer.twitter.com/en/docs/tweets/data-dictionary/overview/intro-to-tweet-json}
containing the tweet text and other important information about the specific tweet. 

In this work, we have used the tweepy\footnote{\url{https://github.com/tweepy/tweepy}} Python library for accessing the Twitter APIs. Tweets related and unrelated to IoT vulnerabilities have been collected from specific Twitter users using the \textit{Get Tweet timelines API}. For the related tweets, we searched Twitter for users with tweets concerning new or existing IoT vulnerabilities; for instance, user ``CVEnew'' that is the official Twitter account maintained by the CVE team to notify the community of new CVE IDs. We also searched for users (e.g., ``IOTASupport'', ``IoTCommunity'', etc.), whose tweets concerned IoT devices in general, new trends, updates, etc.; these tweets did not announce new or existing vulnerabilities, thus they were used as the unrelated tweets. Apart from that, related and unrelated tweets were gathered using the \textit{Twitter Streaming API} by monitoring tweets that contain specific keywords such as ``IoT'', ``iotvulnerability'', etc.

The testing dataset has been assessed by reviewers (postgraduate students in the area of cyber-security) and each collected tweet has been characterized as related or unrelated to IoT vulnerabilities; this decision was used later as the ground truth for the testing dataset in the classification process. Notice that the testing dataset is not balanced, since most entries were tweets unrelated to IoT vulnerabilities. The skewness of the dataset reflects to the real world, as the majority of posted IoT-related tweets concern IoT devices' architecture, updates, use cases, etc., but not IoT vulnerabilities. In the current dataset~\cite{myrepo} there were 4,200 tweets, with 3,953 tweets unrelated and 247 related to IoT vulnerabilities. Table~\ref{table:related-not-related-tweets-test} presents some examples of related and unrelated to IoT vulnerabilities tweets.

Finally notice that, prior to classification, all datasets (training, validation and testing) have been preprocessed as follows: HTML decoding, URL link removal, word tokenization, stopword and special character removal.

%% file: sec4/sec4.tex
\section{Tweet classification using ML Algorithms} \label{sec:classification}

\begin{table*} [t]
    \centering
    \caption{Example False Negative and False Positive Tweets}\label{table:fn-fp-tweets}
    \vspace{-1mm}
    \begin{tabularx}{\textwidth}{XX}
        \toprule
        \textbf{\centering Related to IoT vulnerabilities}  & \textbf{Unrelated to IoT vulnerabilities} \\
        \midrule 
        \textit{BLOG: Recently, a cyber threat actor audaciously cracked into the systems of a Florida water treatment plant and ordered \ldots} & \textit{CVE-2020-27539 Heap overflow with full parsing of HTTP response in Rostelecom CS-C2SHW 5.0.082.1. AgentUpdater service has a self-written HTTP parser and builder. HTTP parser has a heap buffer overflow (OOB write). In default configuration camera parses \ldots} \\ 
        \midrule
        \textit{Misconfigured Baby Monitors Allow Unauthorized Viewing \#CloudSecurity \#IoT \#Vulnerabilities \#InfoSec \#MobileSecurity \ldots}  &   \textit{Flaws in Wireless Mice and Keyboards Let Hackers Type on Your PC.}\\
        \bottomrule
    \end{tabularx}
\end{table*}

\begin{table} [t]
    \centering
     \caption{Performance comparison of classification models}\label{table:test_classification_metrics}
     \vspace{-1mm}
    \begin{tabular}{@{}llccccc@{}}
        \toprule
        \scriptsize{\textbf{Classifier}} &\scriptsize{\textbf{Training}} &\scriptsize{\textbf{F1-score}} &\scriptsize{\textbf{Precision}} &\scriptsize{\textbf{Recall}} &\scriptsize{\textbf{Accuracy}}\\
         &\scriptsize{\textbf{Dataset}}& & & & & \\
        \midrule
        \scriptsize{\textbf{Random Forest}} &\scriptsize{\textbf{No CVE}}  &\scriptsize{\textbf{0.7842}} &\scriptsize{\textbf{0.7265}} &\scriptsize{0.9090} &\scriptsize{\textbf{0.9323}}\\
         &\scriptsize{CVE} &\scriptsize{0.7765} &\scriptsize{0.7175} &\scriptsize{\textbf{0.9157}} &\scriptsize{0.9271}\\
        \midrule
         \scriptsize{Decision Tree} &\scriptsize{No CVE} &\scriptsize{0.7283} &\scriptsize{0.6773} &\scriptsize{0.8736} &\scriptsize{0.9050}\\
          &\scriptsize{CVE} &\scriptsize{0.7224} &\scriptsize{0.6720} &\scriptsize{0.8784} &\scriptsize{0.8997}\\
        \midrule
        \scriptsize{Multinomial NB} & \scriptsize{No CVE} &\scriptsize{0.4716} &\scriptsize{0.5533} &\scriptsize{0.7376} &\scriptsize{0.6026}\\
         &\scriptsize{CVE} &\scriptsize{0.4712} &\scriptsize{0.5531} &\scriptsize{0.7372} &\scriptsize{0.6019}\\
        \midrule
        \scriptsize{Logistic Regression} &\scriptsize{No CVE} &\scriptsize{0.5824} &\scriptsize{0.5860} &\scriptsize{0.8074} &\scriptsize{0.7697}\\
         & \scriptsize{CVE} &\scriptsize{0.5424} &\scriptsize{0.5712} &\scriptsize{0.7836} &\scriptsize{0.7142}\\
        \midrule
        \scriptsize{SVM} &\scriptsize{No CVE} &\scriptsize{0.6049} &\scriptsize{0.5974} &\scriptsize{0.8297} &\scriptsize{0.7938}\\
         & \scriptsize{CVE} &\scriptsize{0.5625} &\scriptsize{0.5798} &\scriptsize{0.8063} &\scriptsize{0.7390}\\
        \midrule
        \scriptsize{Knn} & \scriptsize{No CVE} &\scriptsize{0.5080} &\scriptsize{0.5569} &\scriptsize{0.7392} &\scriptsize{0.6700}\\
         &\scriptsize{CVE} &\scriptsize{0.4525} &\scriptsize{0.5432} &\scriptsize{0.6938} &\scriptsize{0.5809}\\
        \bottomrule
    \end{tabular}
    \vspace{-4mm}
\end{table}

The proposed monitoring system consists of two stages: (i) the data acquisition phase described in the previous section and (ii) the trained classification model that classifies the collected tweets using ML algorithms.
In this work, we aim for binary classification as we want to classify tweets into two distinct classes, namely related and unrelated to IoT vulnerabilities. 
We tested several traditional ML models to figure out which best fits to our case, namely, Logistic Regression, Multinomial Naive Bayes, Decision Tree Classifier, K-Nearest Neighbors Classifier, Support Vector Machines and Random Forest Classifier; the best performing algorithm would then be used in the monitoring system as the classification model. All the above classification models have been trained and evaluated with the same datasets (i.e., both versions of the training dataset).

One critical issue in ML algorithms is the hyper-parameter optimization, which is a time-consuming process. In this work, we used the classifiers from the Scikit-learn Python library~\cite{scikit-learn}; GridSearch and 10-fold cross-validation were used to evaluate the accuracy of the models and for tuning the hyper-parameters; all the experiments were set-up on Google Colab~\cite{colab}.

\textbf{Logistic Regression}~\cite{LogisticRegression} is a classification method, 
which measures the relationship between the categorical-dependent variable and one or more explanatory variables, by estimating probabilities using a logistic (sigmoid) function. For this work, \texttt{penalty} was set to 12, \texttt{C} was set to 10, \texttt{solver} was set to ``newton-cg'', and all other parameters were set to default values.

\noindent\textbf{Multinomial Naive Bayes}\cite{Bayes, naive-bayes} is a classification method that is based on Bayes’ Theorem. The assumption of this classifier is that the presence of a feature in a class is unrelated to the presence of any other feature and that all these properties independently contribute to the final probability. Any vector that represents a text contains information about the probabilities of appearance of the words of the text within the texts of a given class, so the algorithm can compute the likelihood of that text to belong in a specific class. This classification method is suitable for classification with discrete features like word counts on text classification. In this work, \texttt{alpha} parameter was set to 0.1, \texttt{class\_prior} was set to None, and \texttt{fit\_prior} was set to True.

\noindent\textbf {Decision Trees}\cite{tree} can be used for classification and regression problems and can handle categorical and numerical data. To do so, a tree that consists of decision nodes and leaf nodes is created; a decision node can have one or more branches, whereas a leaf node is the node that represents a classification label or a decision. The root node of the tree is the node that corresponds to the best predictor. The basic disadvantage of that method is that the tree usually overfits to its training dataset. While implementing this method, we used ``Gini'' for the \texttt{criterion} parameter, \texttt{min\_samples\_leaf} was set to 5, and all other parameters were set to default values. 

\noindent\textbf{K-Nearest Neighbor}\cite{knn} is a classification algorithm which puts labels on each data point by looking at the ``K'' labelled neighboring points closest to it and by assigning the top-most label of the neighbors; ``K'' is the number of neighbors that will be checked for each data point. In our model, number of \texttt{neighbours} was set to 5, \texttt{weights} were uniformly distributed, ``euclidean distance'' was used as \texttt{metric}, and all other parameters were set to default. 

\noindent\textbf{Support Vector Machines}\cite{svm} can be used for both regression and classification problems. The objective of this algorithm is to find a hyperplane in an N-dimensional space, where N is the number of features that distinctly classify the data points. Since there are many possible hyperplanes that may classify the data points, the plane that has the maximum margin (distance between data points of both classes) must be land. In this work, \texttt{C} parameter was set to  1.0, \texttt{gamma} parameter was set to ``scale'', sigmoid used as \texttt{kernel} function, and all other parameters were set to default values. 

\noindent\textbf{Random Forest}\cite{RandomForestClassifier} is an ensemble learning method, which can be used for classification and regression problems, and belongs to the decision tree family of algorithms. In more details, it uses a group of decision trees, where each tree is created with a different chosen subset of the training data and with a chosen subset of the features at every node; the choice of the training subset and features is random. One main advantage of this classifier is that it overcomes the overfitting issue on the training dataset, which in turn tends to limit the performance of the decision trees. The main reason why this model does not suffer from the overfitting problem is that it takes the average of all the predictions that cancels out the biases. Also, this classification model is able to perform well in the case of missing data by  using median values to replace continuous variables or by computing the proximity-weighted average of the missing values. In this work, the number of \texttt{estimators} was set to 300, \texttt{max\_features} was set to 0.75, and all other parameters were set to default values.

To fully evaluate the effectiveness of the classification models, both precision and recall should be examined~\cite{schutze2008introduction}. Precision is calculated as the fraction of the truly related tweets among the ones classified as related, while recall is calculated as the fraction of the related tweets that were discovered.
High values of precision means that false positives (FP) are minimized, whereas high recall values mean that tweets are correctly classified.  
Table~\ref{table:test_classification_metrics} presents the performance measures for all classification models when using either version of the training dataset.

As shown by the resulting values, the classification models performed better when trained with no CVE-tagged data. Concerning precision, all the classifiers achieved lower values compared to the Random Forest algorithm; Random Forest resulted in the least FPs. Additionally, all models used, except Random Forest, achieved recall values between 0.7 and 0.9, which means that all classification algorithms classified correctly most of the related tweets for both versions of the training dataset. However, Random Forest still performed better than its competitors, as it resulted in the highest recall value (i.e., above 0.9). The same results are observed in accuracy metric where Random Forest has the highest accuracy metric above 0.9. The problem with using accuracy as the main performance metric is that it does not do well when there is a severe class imbalance for that reason we focus on the F1-score, which is a weighted harmonic mean of recall and precision. It is obvious that Random Forest classifier, with F1-score above 0.78, achieved the highest performance, whereas all other classification models achieved F1-score lower than 0.73. Thereby, Random Forest outperformed its competitors for both versions of the training dataset.

Looking closer, the testing dataset consisted of 4,200 tweets, 23.47\% of which were retweets. Random Forest managed to classify correctly 3,916 tweets (TN:3,698, TP:218) out of 4,200 total tweets when trained with no CVE-tagged data, whereas it classified correctly 3,894 tweets (TN:3,671, TP:223) when trained with CVE-tagged data. 
In addition, FPs were slightly less in the case that no CVE-tagged data were used for the training of the model. On the contrary, false negatives (FN) were five more (29 vs. 24) in this case; notice though that only two tweets were actually misclassified as three of them were retweets.
However, FN tweets are ambiguous enough to also be wrongly classified by human experts; Table~\ref{table:fn-fp-tweets} presents some example FN and FP tweets.

Considering all the above, we used the no CVE-tagged trained Random Forest model in our monitoring system.

%% file: sec5/sec5.tex
%\vspace{-1mm}
\section{Monitoring System}\label{sec:monitoring}

The Social Media Monitoring system focuses on real-time event detection from Twitter streams using state-of-the-art tools from the data science domain to automatically classify posts as related or unrelated to IoT vulnerabilities. To verify the performance of our system in a real-world setting, we performed two distinct 30-minute runs of a full monitoring cycle (Twitter Stream API + Classification).  Since the collected tweets were evaluated by human experts, we opted for a short testing period to keep the evaluation manageable; we plan to make more extensive testing in the future. We used 11 keywords, related to the domain of IoT security in the Twitter Stream API. These keywords are: ``IoT'', ``iotvulnerability'', ``exploitIOT'', ``remote attacker'', ``cyberSecurity'', ``cyberAttack'', ``snapdragon'', ``IoTsecurity'', ``CVE'', ``affected device'', ``devices firmware''. For the classification process, we used the Random Forest classifier described above. 

\begin{figure}[t]
\vspace{-6mm}
\centerline{\includegraphics[width=0.54\textwidth]{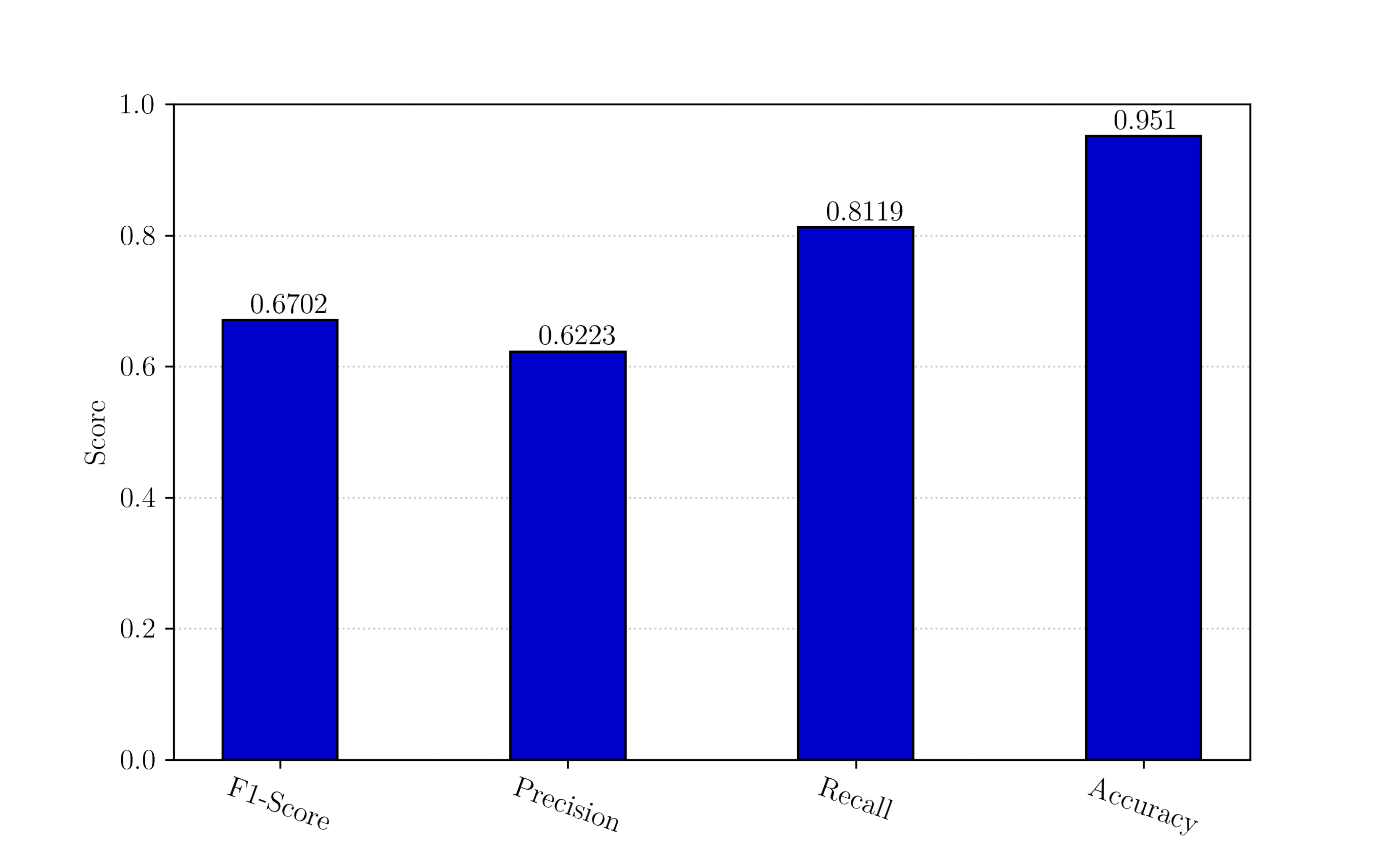}}
\vspace{-1mm}
 \caption{Performance of the Social Media Monitoring System}
\vspace{-4mm}
\label{fig:conf-matrix-monitor}
\end{figure}

The monitoring system acquired 2,126 tweets, out of which 2,007 were identified as unrelated and 119 as related, and harvest rate was equal to 5.9292\%. Upon a close examination, 51\% of the monitoring tweets were retweets, which can be used as an identification metric of the importance of a tweet, (i.e.~the more times a tweet is retweeted the more important it could be). 
The ground truth of the collected tweets was provided by human experts who assessed each collected tweet as related/non-related to IoT vulnerabilities.
Similarly to most real-life classification problems, in this case imbalanced class distribution existed and thus, F1-score is more suitable to evaluate our classification model. The monitoring system achieved 0.67 F1-score, 0.81 precision, 0.62 recall and 0.95 accuracy; all the results are shown in more detail in Figure~\ref{fig:conf-matrix-monitor} and are calculated over both related/unrelated classes. This emphasizes the importance of collecting only relevant tweets as we are dealing with a streaming service, where mis-classification of irrelevant tweets leads to information overload.
 
\begin{comment}
\begin{table} [t]
\centering 
 \caption{Classified Tweets in Numbers}
 \vspace{-2mm}
 \label{table:perf-metrics-monitor}
	\begin{tabular}{l|l|c|c|c}
		\multicolumn{2}{c}{}&\multicolumn{2}{c}{Actual}&\\
		\cline{3-4}
		\multicolumn{2}{c|}{}&Positive&Negative\\
		\cline{2-4}
		\multirow{2}{*}{Predicted}& Positive & 30 & 89 \\
		\cline{2-4}
		& Negative & 15 & 1992 \\
		\cline{2-4}
	\end{tabular}
\vspace{-4mm}
\end{table}
\end{comment}

%% file: sec7/sec7.tex
\section{Conclusions} \label{sec:conclusion}

We proposed, designed, and evaluated a Twitter monitoring system, which aims at the identification of IoT CTI; our design was based on the construction and public release of a new dataset specifically created for the task and on the careful evaluation of several classification alternatives using this dataset. Our approach was evaluated with real-world tweets and model achieved high F1-score, recall, precision and accuracy despite the particularities of the classification task.

Our future research plans involve (i) further investigating the performance of Random Forest for the specific task in comparison to other advanced classification algorithms, (ii) evaluating the classification performance of NNs/DNNs for this task and comparing it to that of the traditional ML algorithms, (iii) extending the monitoring system with a ranking component to take into account different user and tweet metrics (e.g., reliability, popularity, freshness), and (iv) leveraging the correctly identified CTI to actionable insights by employing named entity recognition methods to identify key concepts in the text.